\documentstyle[preprint,aps]{revtex}
\newcommand{\be}{\begin{equation}} \newcommand{\ee}{\end{equation}} 
\newcommand{\bea}{\begin{eqnarray}}\newcommand{\eea}{\end{eqnarray}}

\begin{document}
\draft
\title{{Chiral Solitons in a Current Coupled Schr\"odinger Equation
With Self Interaction  }} 
\author{E. Harikumar, C. Nagaraja Kumar and M. Sivakumar$^{**}$}
\address{School of Physics \\
University of Hyderabad\\
Hyderabad, Andhra Pradesh\\
500 046 INDIA.}
\footnotetext {$^{**}$ Electronic address: 
mssp@uohyd.ernet.in }  
\maketitle

\begin{abstract} 
Recently non-topological chiral soliton solutions were obtained
in a derivatively coupled non-linear Schr\"odinger model in
$1+1$ dimensions.  We extend the analysis to include a more
general self-coupling potential (which includes the previous
cases) and find chiral soliton solutions.  Interestingly even
the magnitude of the velocity is found to be fixed.  Energy and
$U(1)$ charge associated with this non-topological chiral
solitons  are also obtained.

\end{abstract}
\pacs{11.10Lm, 03.50Kk}
\draft
\newpage
The study of solitons, localized travelling solutions with a
finite energy density in non-linear scalar field theories, has a
long history \cite{RR}. In recent times $2+1$ dimensional
theory with matter fields coupled to gauge fields governed by
Chern-Simons action also is found to have soliton solutions,
which is of relevance in Quantum Hall Effect \cite{RJ}.

In a recent work, a derivatively coupled scale invariant
non-linear Schr\"ondinger equation in $1+1$ dimensions, obtained
partially by the dimensional reduction of a $2+1$ dimensional
theory, was shown to have a novel, soliton solution: soliton
exists only for a fixed sign of velocity, but for a range of
magnitude \cite{AU}
.Such solitons are likely to have applications in
one-dimensional system like quantum wires and in the description
of chiral waves which are travelling edge states, in the quantum
Hall effect \cite{XG}. Apart from its potential application, it
is of intrinsic interest to see if the known non-linear equation
can be modified to admit solitons, which travel only
unidirectionally. Such studies have been carried out recently
for generalised KdV and other non-linear equations \cite{DB}
and also for the multicomponent non-linear Schr\"ondiger
equation\cite{MH}.  In this letter, we extend the analysis of
\cite{AU}, which introduced the current coupled non-linear
Schr\"ondinger equation, to include a more general self-coupling
and study the soliton solution of the model.

 The non-linear Schr\"odinger Equation in $1+1$ dimension, with
cubic non-linearity, given by
\be
i{\hbar} {\partial}_t \psi = -{\frac{{\hbar}^2} {2m}}
{\partial}^2_{x} \psi - g (\psi^* \psi)\psi
\ee
\noindent is well studied and soliton solutions are constructed. This model
is found to be completely integrable.  Recently a new non-linear
Schr\"odinger Equation was constructed, which has non-linearity
due to current coupling, rather than charge density coupling in
the equation:

\be
i{\hbar} \partial_{t} \psi = -{\frac{{\hbar}^2} {2m}}
{\partial}^2_{x} \psi - \lambda j(x,t) \psi(x,t)
\ee
\noindent 
where $j=\frac{\hbar}{m} Im(\psi^* \partial_x \psi)$ is the 
current 
density.  This derivative non-linear Schr\"odinger (DNLS)
equation was found to have soliton solutions which are (i)
chiral - i.e., having only a fixed sign of velocity, (ii)
velocity can not be arbitrarly reduced -i.e., absence of
Galilean invariance. Note that the equation (2) belongs to 
same class of the DNLS of Kaup and Newell \cite{DJ} which reads
\be
i \partial_{t} \psi = -~{\partial}^2_{x} \psi ~~{\pm}~i~ 
{\partial}_{x} (\psi^* {\psi}^2).
\ee

Eqn(2) can not be dircetly obtained from a local Lagrangian.
Instead, as shown in\cite{AU}, the following Lagrangian density
provides an equation of motion equivalent to that of (2), by a
suitable redefinition of the field, as indicated below.

\be
{\cal L} = i{\hbar} \Phi^* {\partial_t \Phi} - {\frac
{{\hbar}^2} {2m}}
\left | (\partial_x +i{\lambda\over 2} \rho)\Phi\right|^2
\ee
\noindent where $\rho = \Phi^*\Phi$.

The equation of motion following from the above Lagrangian is,
\be
i{\hbar}{\partial_t \Phi} = - {\frac {{\hbar}^2} {2m}} {\left
(\partial_x +i{\lambda\over 2} \rho\right)}^2\Phi +
\frac{\hbar}{2}
\lambda J \Phi,
\ee
\noindent where 
\be J=\frac{\hbar}{m} Im \left(\Phi^*(\partial_x +i
\frac{\lambda}{2}\rho) \Phi\right) ,
\ee
\noindent  obeys the continuity equation
\be
 \partial_t ({{\phi}^*{\phi}})~+~\partial_x J(x)~=~0.
\ee
\noindent By redefining the field $\Phi$ by
\be
\Phi = exp~~ ({i}\frac{\lambda}{2} {\int {dy} \rho(y,t)})\psi,
\ee

\noindent eqn(2) is obtained after using continuity equation(7) and eqn(6)
for $J$.

Incidentally, the  Lagrangian density (4) is related to $2+1$
dimensional non-relativistic field coupled to an $U(1)$ gauge
field, whose kinetic term is the Chern-Simon term, by
dimensional reduction. In the ensuing $1+1$ dimensional
theories, for $A_3 =B(x,t)$ field, which is non-propagating, a
kinetic term of the form $ \frac{dB}{dt}\frac{dB}{dx}$ is added
by hand.  By using Hamiltonian reduction, (and by a phase
redefinition of $\psi$) $A_{\mu}(x,t)$ and $B(x,t)$ can be
eliminated resulting in (4).

In \cite{AU}, an interesting non-topological soliton solution
was discovered for this novel, derivative coupled non-linear
Schr\"odinger equation, having chiral motion in a global rest
frame, when $V({\phi}^*{\phi})$ is absent and also when
$V({\phi}^*{\phi})$ is repulsive cubic. It should be of interest
to examine whether these type of solutions exist for other
interactions.  In this note, we extend the analysis of
\cite{AU} to include a more general nature of anharmonicities (
potential ) and discuss the nature of the soliton solutions in
this model. The potential we add to the Lagrangian is of the
form
\be
 V({\rho}^2)~=~a{{\rho}^{2n+2}}~+~b{{\rho}^{n+2}}~+~c{{\rho}^4},
\ee
\noindent where ${\rho}^2~=~(\phi^* \phi)$ and a, b, and c are
dimensionful coupling constants. The $n=1$ and $n=2$ cases are
included in the \cite{AU}.
\noindent Such anharmonic potential terms has been considered earlier 
in the context of models describing both first and second order
transitions, and topological, nontopological and periodic
solutions were obtained \cite{SN}.

	Interestingly we find, as shown below, except for $n=1$
and $2$, soliton solutions are of a fixed velocity (with fixed
magnitude and sign).  This  has to be contrasted with the result
of \cite{AU}, where only the sign of the velocity has fixed.
Such velocity selection is also present  in a modified Sine
-Gordon theory \cite{MP}. We also calculate the $U(1)$ charge
and energy of these soliton solutions.

The equations of motion following from the Lagrangian (4), with
the potential (9) added is,
\be
i\hbar {\partial}_t{\phi}~=~ - {\frac {\hbar^2} {2m}}{\left [
\partial_x +~{\frac{i\lambda}{2}}{\rho^2}\right]}^{2} \phi~+~
{\frac {\lambda \hbar}{2}}J\phi~+~{V^\prime}\phi,
\ee
\noindent where $V^{\prime}~=~ \frac{dV}{d{\rho}^2}$.
In order to construct the soliton solution  make the ansatz,
\be
\phi~=~ \rho(x~-~vt){e}^{i\theta(x,t)}
\ee
\noindent where $v$ is the velocity. Using (11) in (7) one gets
\be
{\theta}^\prime~=~ \frac{mv}{\hbar}~-\frac{\lambda {\rho}^2}{2}.
\ee

\noindent Using (12)and (11), the equation of motion (10) becomes
\be
\rho^{\prime\prime}~=~\left[\left({\frac {mv}{\hbar}}\right)^2~+
~{\frac{2m\omega_0}{\hbar}}\right] \rho ~+~
{\frac{mv\lambda}{\hbar}}{\rho}^3 ~+~ {\frac{2m}{\hbar}}{\frac{d
V}{d\rho}}.
\ee
\noindent where $\omega_0 ={\partial_{t} \theta}$ and
 $${\frac{d V}{d\rho}}~=~~\left(~a(2n~+~2){\rho}^{2n+1}~+
~b(n~+~2){\rho}^{n+1}~+~4c{\rho}^3\right)$$
\noindent Note for $n=1$ and $n=2$, eqn(13) reduces to that
considered in \cite{AU}.  This non-linear equation admits a
localized solution for $n> 2$,
\be
\rho(x-vt)~=~\left[{\frac{2mb}{\hbar}}{M}^{-2}{\sqrt{1-{\frac{a}{m{b}^2}} 
{M}^2}}~\right]^{-\frac{1}{n}} {\left[{\frac{1}{
{\sqrt{1-{\frac{a}{m{b}^2} {M}^2}
}}}}~+~cosh(Mnx)\right]^{-\frac{1}{n}}}
\ee
\noindent only when the velocity 
\be
v = - ({\frac{8c}{\lambda}}),
\ee
\noindent where 
$$M~=~\left[\left({\frac{m v}{\hbar}}\right)^2~
+~{\frac{2m\omega_0}{\hbar}}\right]^{\frac{1}{2}}.$$

Note that sign of $\lambda$ determines the direction of
velocity, implying the chiral nature of the solitons. Chirality
is due to the constraint on $v$ rather than on $v^2$. Note that
the magnitude of the velocity is also fixed.

In the case, when $ n = 1 $ and $2$ the solution $\rho ( x, t )
$ is valid without any restriction on the magnitude of the
velocity.  Chirality nature of the solution still exists for
these cases except for $n=2$  and $ a< 0 $ as shown in
\cite{AU}.

The conserved $U(1)$ charge associated with the soliton solution
is
\bea
N
=\!\!\!\!\!\!\!\!\!\!\!\!\!\!\!\!\!
\!\!\!\!\!\!\!\!\!\!\!&\int{dx}{(\phi^*\phi)}\nonumber\\
=\!\!\!\!\!\!\!\!&\frac{1}{M n(\alpha\beta)^{\frac{2}{n}}}\int
{\frac {dy}{\left({\frac{1}{\beta}}+coshy
\right)^{\frac{2}{n}}}}\nonumber\\ =&{\frac{1}{M
n{(\alpha\beta)}^{\frac{2}{n}}}} {\frac {\beta}
{\sqrt{1-{\beta}^2}}} Q_{({\frac{2-n}{n})}}
({\frac{1}{\sqrt{1-{\beta}^2}}}),
\eea
\noindent where $$\alpha= {\frac{2 m b}{\hbar}} {M}^{-2},$$
$$\beta~=~{\sqrt{1~-~{\frac{a{M}^2}{m {b}^2}}}}$$ and
$Q_{\nu}(k)$ is Legendre function of second kind \cite{IS}.

Energy associated with this soliton configuration is
\be
{\cal E}~=~\int{dx}{\cal H},
\ee
\noindent where
\be
{\cal H}~=~ \frac{{\hbar}^2}{2m} {|D_x\phi|}^2~+~V({\rho}^2).
\ee
\noindent is  the Hamiltonian density.
\noindent Substituting the ansatz (11), and using (13) in (18) we get 

\be
{\cal E}
=\int{dx}{\frac{{\hbar}^2}{2m}}\left[~(\rho^{\prime})^2~+~
\left({\frac{mv}{\hbar}}\right)^2 {\rho}^2~+
~ {\frac{2m}{{\hbar}^2}} V({\rho}^2)~\right]\nonumber\\
\ee
Now substituting the solution (14)
\bea {\cal E} &=\left[ M^2~+~{\left(mv\over{\hbar}\right)}^2\right] N -
{1\over{n {(\alpha\beta)}^{({{n+2}\over n})}}}\left[ 2M\alpha -
{2mb\over{{\hbar}^2 M}}\right]
{\left[{\beta}^2\over{1+{\beta}^2}\right]}^{-{({{n+2}\over
2n})}} Q_{2\over n}{({1\over{\sqrt{1+{\beta}^2}}})}\nonumber\\
+\!\!\!\!\!\!  & {1\over{n {(\alpha\beta)}^{({{2n+2}\over n})}}}
\left[ M{\alpha}^2 {(1-{\beta}^2)}+{2 m a \over{{\hbar}^2 M}}\right]
{\left[{\beta}^2\over{1+{\beta}^2}\right]}^{-({{2n+2}\over 2n})}
Q_{({n+2\over n})}{({1\over{\sqrt{1+{\beta}^2}}})}\nonumber\\
\!\!\!\!\!\!\!\!\!\!\!\!\!\!\!\!\!\!\!\!\!\!\!\!\!\!\!\!\!\!\!\!\!\!\!\!\!\!
\!&{2mc\over{{\hbar}^2 Mn{(\alpha\beta)^{4\over n}}}}
{\left[{\beta}^2\over{1+{\beta}^2}\right]}^{-2\over n}
Q_{({4-n\over n})}{({1\over{\sqrt{1+{\beta}^2}}})}
\eea
\noindent Here $Q_{\nu}^{\mu}$ are the associated Legendre functions of
second kind and $Q_{\nu}^{0}~=~Q_{\nu}$.  Interestingly for both
attractive and repulsive cases of highest anharmonicity $( a > 0
~~~{\rm and}~~~ a < 0 )$, finite energy chiral soliton solution
exists.  The study of interesting cases of topological soliton
and soliton with non-trivial boundary conditions ( i.e., with
non-zero density at spatial infinity ) is in progress.

\noindent {\bf Acknowledgements}:
One of us (EH) thanks U.G.C., India for support through J.R.F
scheme.  (CNK) thanks CSIR, India, for support through S.R.A.,
scheme. MS work is partially supported by DST project.


\end{document}